\documentclass[conference]{IEEEtran}
\IEEEoverridecommandlockouts

\usepackage[T1]{fontenc}
\usepackage[compress]{cite}
\usepackage{paralist}
\usepackage{newunicodechar}
\usepackage{wrapfig}
\usepackage{tabularx}
\usepackage{tablefootnote}
\usepackage{multirow}
\usepackage{booktabs}
\usepackage{subcaption}
\usepackage{xcolor}
\usepackage{colortbl}
\usepackage{fontawesome5}
\usepackage{tikz}
\usepackage{balance}
\usepackage[spaces,hyphens]{xurl}
\usepackage[colorlinks,allcolors=blue]{hyperref}

\usepackage[noframe]{showframe}
\usepackage{framed}
\usepackage{lipsum}
\renewenvironment{shaded}{%
  \MakeFramed{\advance\hsize-\width \FrameRestore\FrameRestore}}%
 {\endMakeFramed}
\definecolor{shadecolor}{gray}{0.85}

\newcommand{\myparagraph}[1]{\vspace{.5em}\noindent\textbf{#1.}\ }
\renewcommand\subsubsection[1]{\myparagraph{#1}}

\newcommand\figureCont[0]{
\begin{wrapfigure}{r}{0.28\textwidth}
  \begin{center}
     \includegraphics[trim=1cm 1.25cm 1cm 1.75cm, width=0.28\textwidth]{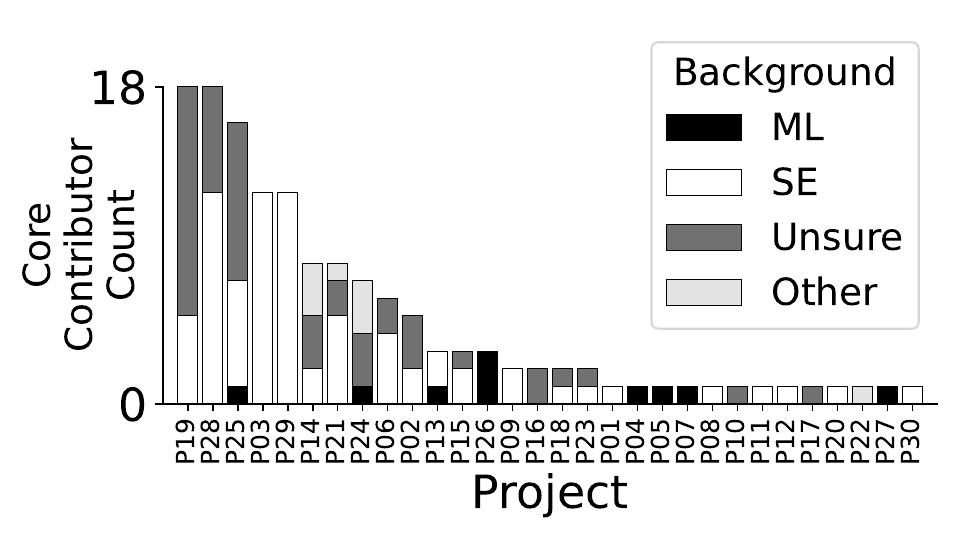}
  \end{center}
\end{wrapfigure}
}
\newcommand\figureContCode[0]{
\begin{wrapfigure}{r}{0.275\textwidth}
  \begin{center}
    \includegraphics[trim=0.25cm 1.5cm 1cm 1.5cm, width=0.275\textwidth]{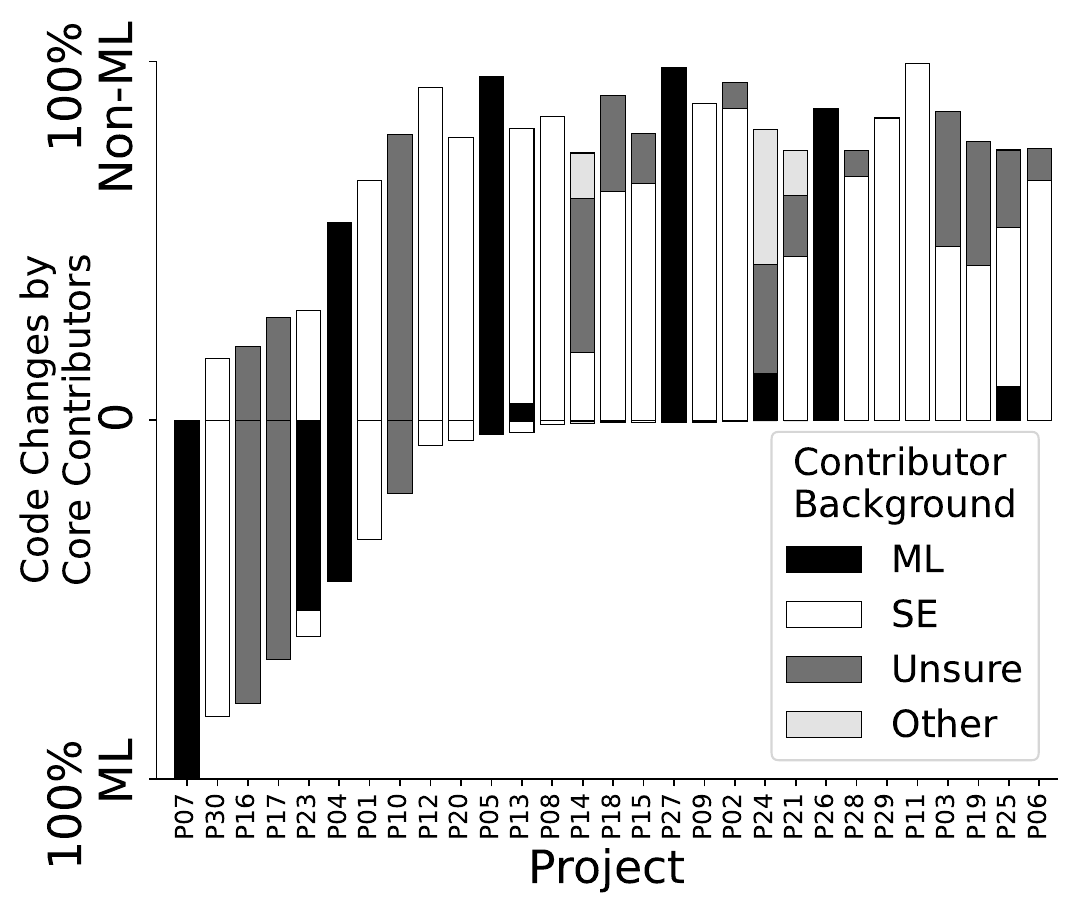}
  \end{center}
\end{wrapfigure}
}

\newcommand\figureModularity[0]{
\begin{wrapfigure}{r}{0.25\textwidth}
  \begin{center}
    \includegraphics[trim=0cm 1cm 1cm 1cm, width=0.25\textwidth]{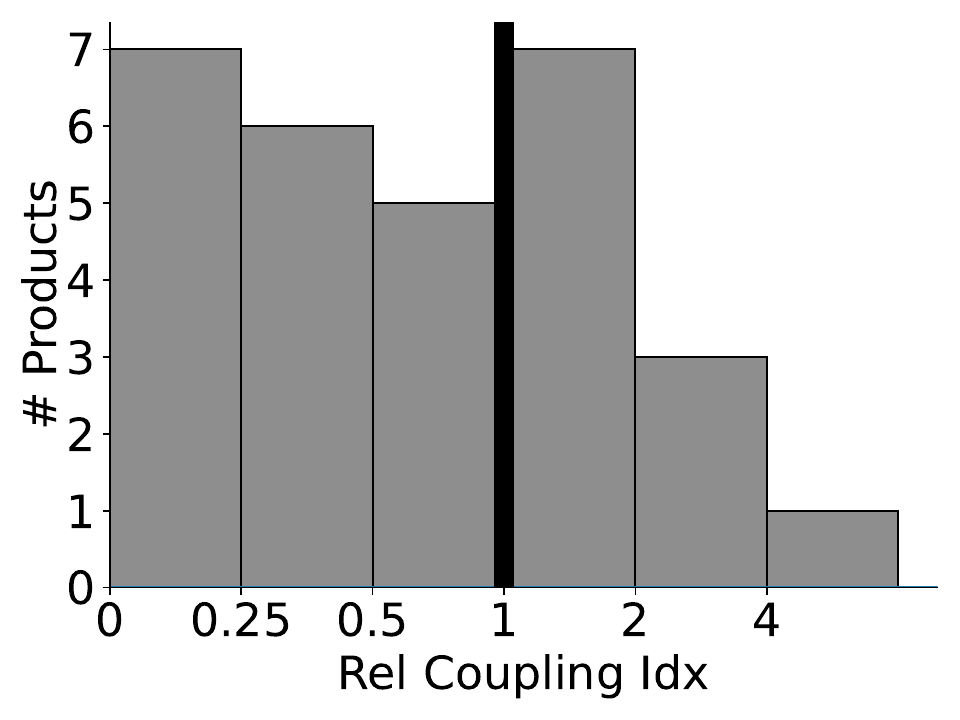}
  \end{center}
\end{wrapfigure}
}

\newcommand\figureSankeyA[0]{
\begin{wrapfigure}{r}{0.2\textwidth}
  \begin{center}
    \includegraphics[trim=0cm 0.5cm 0cm 0.5cm, width=0.2\textwidth]{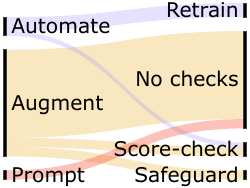}
  \end{center}
\end{wrapfigure}
}
\newcommand\figureSankeyB[0]{
\begin{wrapfigure}{r}{0.2\textwidth}
  \begin{center}
    \includegraphics[trim=0cm 0.5cm 0cm 0.5cm, width=0.2\textwidth]{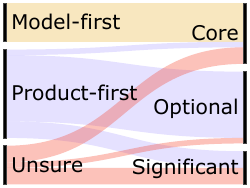}
  \end{center}
\end{wrapfigure}
}

\newcommand\tableA[0]{
\begin{table*}[t]
\caption{Number of Retrieved Projects after Each Step}
\label{numbers}
\centering
\begin{tabularx}{0.75\linewidth}
{@{}l@{\hskip7pt}r@{\hskip7pt}rr@{\hskip7pt}r@{\hskip7pt}r@{\hskip7pt}rr@{\hskip7pt}r@{\hskip7pt}r@{\hskip7pt}r}
\cmidrule{1-11}
\multicolumn{1}{c}{\multirow{2}{*}{\textbf{}}} & \multicolumn{2}{c}{\textbf{Mobile App}} & \multicolumn{4}{c}{\textbf{Desktop App}} & \multicolumn{3}{c}{\textbf{Web App}} & \multicolumn{1}{c}{\multirow{2}{*}{\textbf{Total}}} \\\cmidrule(lr){2-3}\cmidrule(lr){4-7}\cmidrule(lr){8-10}
\multicolumn{1}{c}{} &
  \multicolumn{1}{c}{\textbf{android}} &
  \multicolumn{1}{c}{\textbf{iOS}} &
  \multicolumn{1}{c}{\textbf{js}} &
  \multicolumn{1}{c}{\textbf{py}} &
  \multicolumn{1}{c}{\textbf{java}} &
  \multicolumn{1}{c}{\textbf{C\#}} &
  \multicolumn{1}{c}{\textbf{js}} &
  \multicolumn{1}{c}{\textbf{py}} &
  \multicolumn{1}{c}{\textbf{java}} & 
\multicolumn{1}{c}{}\\
\cmidrule{1-11}
\textbf{GitHub Search}                                & 12,044          & 10,969         & 72,793  & 67,626  & 55,892 & 15,267 & 72,793   & 67,626   & 55,892   & 430,902\\
\textbf{Initial Filter}                          & 3,358           & 4,055          & 83,777  & 36,396  & 22,802 & 7,145  & 83,777   & 36,396   & 22,802   & 300,508 \\
\textbf{Product Filter}                              & 2,296           & 2,801          & 1,100   & 1,663   & 1,909  & 2,590  & 3,025    & 8,747    & 2,255    & 26,386\\
\textbf{Manual Check*}                            & 33 & 14         & 19 & 43 & 17 & 12          & 42 & 104 & 5 & \textbf{**262}\\
\cmidrule{1-11}
\multicolumn{11}{r}{*Based on the ML clue counts, we inspected around 4k projects manually. **Removing duplicates.}                                                             
\\
\end{tabularx}
\end{table*}
}

\newcommand\tableB[0]{
\begin{table}[t]
\caption{Sample ML Products for Analysis, from the Curated Dataset of 262 ML Products: Mobile (P1-P10), Desktop (P11-P20), and Web Applications (P21-P30)}
\label{samples}
\scriptsize
\begin{tabularx}{\linewidth}{l@{\hskip3pt}p{1.9cm}@{\hskip3pt}p{3.75cm}@{\hskip3pt}r@{\hskip3pt}r@{\hskip3pt}r}
\toprule
\textbf{ID}   & \textbf{Name} & \textbf{Descrption} & \textbf{Star} & \textbf{Cont.} & \textbf{U/D*} \\\midrule
P1  & \href{https://github.com/renard314/textfairy}{Text Fairy}                   & OCR scanner app & 751        & 5 & 10M+        \\
P2  & \href{https://github.com/inaturalist/SeekReactNative}{Seek by iNaturalist}         &  App to identify plants and animals & 92 & 8 & 1M+\\
P3  & \href{https://github.com/Catrobat/Catroid}{Pocket Code}                  & App for learning programming  & 92        & 8 & 7M+\\
P4  & \href{https://github.com/longpth/ESP32CamAI}{ESP32 AI Camera}                    &  ESP32-CAM processing AI tasks & 82 &        1 & 1K+\\
P5  & \href{https://github.com/elblogbruno/NotionAI-MyMind}{NotionAI MyMind}           & App to store and search for items  & 182 &        2 & 1K+\\
P6  & \href{https://github.com/organicmaps/organicmaps}{Organic Maps}               &  Offline map app  & 4023 & 226 & 500K+\\
P7  & \href{https://github.com/prabhakar267/vertikin}{VertiKin}                 &  E-commerce app  & 74 & 5 & N/A\\
P8  & \href{https://github.com/florisboard/florisboard}{FlorisBoard}               &  Android keyboard  & 3503 & 76 & N/A \\
P9  & \href{https://github.com/samuelclay/NewsBlur}{NewsBlur}                   &  Personal news reader  & 6123 & 83 & 50K+\\
P10 & \href{https://github.com/nex3z/tflite-mnist-android}{TfLite MNIST}            &  Handwritten digits classification  & 214 & 1 & N/A\\
P11 & \href{https://github.com/Unidata/awips2}{AWIPS}                  & Advanced weather processing system  & 129 & 6 & 97K/mo \\
P12 & \href{https://github.com/Audiveris/audiveris}{Audiveris}                   & Optical musical recognition app  & 932 &         16 & 8.7K/mo\\
P13 & \href{https://github.com/ICIJ/datashare}{Datashare}                        &  Doc. analysis software for journalists  & 438 & 15 & 1.2K/mo \\
P14 & \href{https://github.com/Capnode/Algoloop}{Algoloop}                      & Algorithmic trading application  & 67 & 160 & N/A\\
P15 & \href{https://github.com/SubtitleEdit/subtitleedit}{Subtitle Edit}             & Editor for video subtitles  & 4407 & 86 & 293K/mo \\
P16 & \href{https://github.com/iperov/DeepFaceLab}{DeepFaceLab}                    &  Software for creating deepfakes  & 35566 &  19 & N/A \\
P17 & \href{https://github.com/deepfakes/faceswap}{Faceswap}                    & Software for creating deepfakes  & 42623         & 80 & 297K/mo \\
P18 & \href{https://github.com/akasolace/HO}{HO}                          &  Helper for Hattrick football manager  & 138 & 12 & 14M+ \\
P19 & \href{https://github.com/bigbluebutton/bigbluebutton}{BigBlueButton}            & Web conferencing system  & 7710 & 181 & 88K/mo \\
P20 & \href{https://github.com/LingDong-/PoseOSC}{PoseOSC}                     & Realtime human pose estimation  & 63 & 2 & N/A \\
P21 & \href{https://github.com/OpenBB-finance/OpenBBTerminal}{OpenBB Terminal}         & Investment research software  & 17481 & 136 & 94K/mo \\
P22 & \href{https://github.com/jgagneastro/coffeegrindsize}{Coffee Grind Size}           &  Coffee particle analyzer  & 402 & 1 & N/A \\
P23 & \href{https://github.com/ritwik12/Celestial-bodies-detection}{Celestial Detection}   &  Classifier of celestial bodies  & 69 & 20 & N/A \\
P24 & \href{https://github.com/electricitymap/electricitymap-contrib}{Electricity Maps} &  Greenhouse gas intensity visualizer & 2566 & 268 & 3M+ \\
P25 & \href{https://github.com/galaxyproject/galaxy}{Galaxy}                  &  Data intensive science for everyone  & 1021 & 255 & 187K/mo \\
P26 & \href{https://github.com/SanPen/GridCal}{GridCal}                        &  Power systems planning software  & 293 & 14 & N/A \\
P27 & \href{https://github.com/castorini/honkling}{Honkling}                    & Keyword spotting system  & 63 & 5 & 1.2K/mo \\
P28 & \href{https://github.com/jitsi/jitsi-meet}{Jitsi Meet}                      &  App for video conferencing  & 18813 & 374 & 10M+ \\
P29 & \href{https://github.com/code-dot-org/code-dot-org}{Code.org}             &  Professional learning program for CS  & 712 & 132 & 82M+ \\
P30 & \href{https://github.com/chonyy/AI-basketball-analysis}{Basketball Analys.}         &  Analyze basketball shooting pose  & 781 & 4 & N/A \\\bottomrule
\multicolumn{6}{p{8.2cm}}{*Users/Downloads: There is no reliable way to calculate the number of users; we report them using multiple ways if available, such as downloads in google play-store, self-reported on website, or website traffic tracker (\url{similarweb.com}) to count average monthly users (in ‘value/mo’ format)}

\end{tabularx}
\end{table}
}

\let\oldbibliography\thebibliography
\renewcommand{\thebibliography}[1]{%
  \oldbibliography{#1}%
  \setlength{\itemsep}{0pt}%
}

\begin{document}
\title{The Product Beyond the Model -- An Empirical Study of Repositories of Open-Source ML Products}

\author{\IEEEauthorblockN{Nadia Nahar\IEEEauthorrefmark{1}\IEEEauthorrefmark{2}, Haoran Zhang\IEEEauthorrefmark{2}, Grace Lewis\IEEEauthorrefmark{3}, Shurui Zhou\IEEEauthorrefmark{4}, Christian Kästner\IEEEauthorrefmark{2}}
\IEEEauthorblockA{\IEEEauthorrefmark{2}Carnegie Mellon University, \IEEEauthorrefmark{3}Carnegie Mellon Software Engineering Institute, \IEEEauthorrefmark{4}University of Toronto\\
\IEEEauthorrefmark{1}nadian@andrew.cmu.edu
}
}

\maketitle
\begin{abstract}
Machine learning (ML) components are increasingly incorporated into software products for end-users, but developers face challenges in transitioning from ML prototypes to products. Academics have limited access to the source of commercial ML products, hindering research progress to address these challenges. In this study, first and foremost, we contribute a dataset of 262 open-source ML products for end users (not just models), identified among more than half a million ML-related projects on GitHub. Then, we qualitatively and quantitatively analyze 30 open-source ML products to answer six broad research questions about development practices and system architecture. We find that the majority of the ML products in our sample represent more startup-style development than reported in past interview studies. We report 21 findings, including limited involvement of data scientists in many open-source ML products, unusually low modularity between ML and non-ML code, diverse architectural choices on incorporating models into products, and limited prevalence of industry best practices such as model testing, pipeline automation, and monitoring. Additionally, we discuss seven implications of this study on research, development, and education, including the need for tools to assist teams without data scientists, education opportunities, and open-source-specific research for privacy-preserving telemetry.
\end{abstract}
\begin{IEEEkeywords}
Open source dataset, machine learning products, mining software repositories, software engineering for machine learning
\end{IEEEkeywords}

        \section{Introduction}\label{h.g6xxiz90iseg}
With the increasing popularity of machine learning (ML), more and more software products are incorporating ML components to enable various capabilities, such as face recognition in photo-sharing apps. However, incorporating ML into products is not just about developing the model; there is a significant amount of effort to integrate the model into the system, while considering aspects such as system architecture, requirements, user experience (UX) design, safety and security, system testing, and operations \cite{rrC2w,RmKB,t8sQ}. Developers find it challenging to convert ML prototypes into \emph{software products with ML components (ML products)} \cite{kLg8,Nrd9,7mTx,a4ccL}. Unfortunately, researchers rarely have access to the source code of ML products and hence face difficulty in (a) studying challenges in-depth and (b) designing and evaluating interventions (e.g., tools and practices). Currently, academic researchers rely primarily on an outside view gathered through interviews or surveys with industry practitioners \cite{7mTx,Nrd9,rrC2w,YIKzr,a4ccL,J91H}. Some researchers perform research within a company and gain rich access \cite{KZWF3,lxnLR,Y9Vl9,0m6Fy,XWAQz}, but are limited to a single context and can rarely share details. This inability to access and study \emph{ML products} -- not just any ML projects -- poses a significant impediment to advancing research in this field, leading to a wealth of academic literature identifying challenges through professionals' testimonies, but a dearth of research offering scientifically-evaluated solutions or interventions at the intersection of software engineering (SE) and ML.

\tableB

Historically, the field of SE has greatly benefited from open-source software in terms of research, practice, and education. Research on mining open-source software repositories has allowed us to create vast datasets and study aspects that would otherwise be challenging to investigate \cite{Ek3M,Tx4J}, such as effectiveness of continuous integration \cite{FW2m} and pull-request-based development \cite{xt2Q}. Open source allows researchers to test interventions on software artifacts, which has been foundational for many research areas such as program repair and software testing \cite{azR0,2h5i}. Many innovations developed and evaluated on open source are later adopted in industry (e.g., Facebook's adoption of program repair \cite{SsxR}, and Google's adoption of mutation testing \cite{vzrE}). Beyond research and development, access to open source on a larger scale has revolutionized education, offering students and professionals the opportunity to study and illustrate practices in open-source repositories \cite{Uq5a,Hr7T,r2EI}. In the same manner, access to open-source \emph{ML products} could open opportunities for research, education, and technology transfer. To this end, we identify a corpus of such ML products.

Many past studies have tried to study ML projects in open source, but usually (a) only focus on one or two specific examples or (b) use a dataset full of notebooks, research projects, homework solutions, and demos, which are not representative of real-world industrial ML products. While open source is useful for studying ML libraries and notebooks \cite{TBW6,IUr7,P8le}, researchers struggle to find good examples of open-source \emph{ML} \emph{products}: Several papers prominently highlight \emph{FaceSwap} \cite{1n0g} as an active end-user open-source ML product and study it in depth, but it is usually the only clear example of an ML product ever identified or analyzed \cite{Ku84,bGZK}. Several papers \cite{OmEh,dx7s,TUkr} rely on a dataset of 4,500 projects labeled as ``ML applied'' \cite{YDPU}, but closer inspection reveals that most of these projects are research notebooks, tutorial-style projects, and toy projects -- not a promising dataset for those interested in studying ML products.  

We have two goals for this paper, (a) to identify a corpus of ML products in open source, beyond just FaceSwap, and (b) to analyze the curated dataset to answer research questions of interest to the community. The first goal turned out to be surprisingly difficult due to the abundance of open-source ML projects that are not ML products, making the keyword search approach used in related studies ineffective. In response, we designed a tailored pipeline, strategized specifically for finding \emph{ML products}. We identified and manually verified a total of 262 repositories \cite{YDPU}. While this is a smaller corpus than past datasets on ML projects, it provides a considerable number of open-source projects that build an end-user product around a model, have a development history, and are fully transparent, providing opportunities not achievable with interviews, surveys, and even industry collaborations.

Second, we analyze our dataset to answer existing research questions for which open-source software might provide useful insights. Instead of a shallow quantitative analysis of all the ML products, we conducted an in-depth analysis of a 30-product sample (Table \ref{samples}) and reported 21 findings around six research questions related to collaboration, architecture, process, testing, operations, and responsible AI. Among others, our findings reveal (a) often limited visible involvement of data scientists in developing open-source ML products, and a lack of clear boundaries between responsibilities for ML and non-ML code, (b) diverse architectural choices on incorporating often multiple models into products, and (c) rare use of industry best practices, such as pipeline automation, model evaluation, and monitoring. While our findings suggest that open-source ML products often mirror practices reported from startup-style ML products in industry, we also find a wide range of practices and products in this dataset. Our dataset offers ample research and educational opportunities, such as developing tools to assist teams that lack access to data scientists, tools and patterns to address the unexpectedly low modularity between ML and non-ML code, and open-source-specific innovations for privacy-preserving telemetry.

To summarize, the primary contributions of this paper are (a) an open-source dataset of 262 ML products, (b) a novel search strategy to identify the ML products from GitHub, and (c) 21 findings around six broad research questions, characterizing the nature of ML products in the dataset.

\section{Defining ML Products}\label{h.wthobesn76g1}
\label{define}

Throughout the paper, we use the term \emph{ML product} to describe software products for end-users that contain ML components. We explicitly distinguish ML products from other ML-related \emph{projects} and artifacts, such as notebooks and models. Note that terminology in this field is not standardized or consistent, as practitioners and researchers may refer to the libraries that train models (e.g., TensorFlow), the code to train models (e.g., in a notebook), the deployed models (e.g., GPT-3), or the products around those models (e.g., FaceSwap) by names such as ML systems, ML projects, or ML applications. Our notion of \emph{ML product} considers the entire software system including non-ML components, in line with past research that used terms like ML-enabled systems \cite{OqUB,rrC2w}, or ML systems \cite{fC0u,Nrd9}.

During our research, we needed a clear definition for what we consider an \emph{ML product} (especially because we had to classify thousands of repositories). In line with \emph{interactive systems} in human-computer interaction \cite{Hgui,6GPK}, and \emph{products} in the context of product management \cite{4TdX,bIwI}, and after iteratively reviewing hundreds of projects (cf. Sec.~\ref{methContA}), we define \emph{ML product} as follows:

\begin{shaded}\noindent


\textbf{ML Product:} A machine-learning product is a software project (a) \textbf{for end-users} that (b) contains one or more \textbf{machine-learning components}.

To be considered \emph{for end-users}, the project must have a \emph{clear} \emph{purpose} and a \emph{clear target audience}. The purpose can be fun and the audience can be ``everybody.'' The software must be \emph{complete, usable}, \emph{polished, and documented} (e.g., install and usage instructions) to the level typically expected by the target audience. The product needs to use at least one machine-learned model that provides major or minor functionality in the software. The model can be developed from scratch or called using an existing library or API.


\end{shaded}

\noindent
For contrast, we define ML libraries and ML projects:

\textbf{ML Library:} Libraries, frameworks, or APIs that are used to perform ML tasks, such as TensorFlow and DVC\footnote{\url{https://dvc.org/}}.

\textbf{ML Project:} ML Project represents any software project that integrates some form of ML functionality or code. Examples include notebooks, research artifacts associated with a paper, and course homework. All ML products are ML projects, but most ML projects are not ML products.

\textbf{Scoping.} We exclude ML products targeting software developers and data scientists as end users from our corpus, such as code-completion tools. These users have more technical expertise and may interact with products through different interfaces, sometimes blurring the line between product and project. Researchers interested in ML products for developers could repeat our search process with a wider scope.

\section{Existing Research and Limitations }\label{h.ay16egamfahe}
\label{relWork}

\textbf{Building ML products is challenging and requires engineering beyond the model and ML pipeline.} Many researchers have studied the challenges that practitioners face when turning an ML model or prototype into a product. We recently collected challenges from 50 papers that surveyed and interviewed practitioners regarding the software engineering challenges faced when building ML products~\cite{qMjN}. These papers illustrated numerous challenges, such as architectural issues due to lack of modularity in ML code increasing design complexity \cite{swt2,Nrd9} and team collaboration hindered by the absence of necessary skills \cite{SuxN,rrC2w}. While these studies with practitioners provide a high-level understanding of the problems, they often do not offer sufficient details or access to design-specific interventions.

\textbf{Researcher tradeoff between internal and external validity for studying ML products.} For studying ML products (not just models and ML projects), researchers adopted different research designs. Some conducted interviews, e.g., \cite{7mTx,Nrd9,swt2,rrC2w}, while others focused on surveying practitioners at scale, e.g., \cite{O0jVj,a4ccL,J91H}. While these studies provide a broad sense of the challenges (maximizing external validity), they rely on self-reported data without access to artifacts. There are also ethnographic studies \cite{Y9Vl9,0m6Fy}, industrial case studies \cite{XWAQz,GtS4}, and experience reports \cite{KZWF3,lxnLR} -- these can yield a deeper understanding of specific cases (maximizing internal validity), but at the cost of generalizability as they are usually based on a single case (a common tradeoff \cite{I0mg}). Access to an open-source dataset of ML products can provide new opportunities, enabling researchers to validate reported challenges on tangible data, devise solutions, and evaluate interventions across a larger number of cases.

\textbf{Academic interventions exist for models and pipelines where plenty of notebooks and libraries exist in open source to study, but not for \emph{product-level} problems to which academics hardly have access.} Academics can design and evaluate solutions when adequate data is available, such as the millions of notebooks on GitHub \cite{IUr7,P8le}, for example supporting studies and evaluations of interventions on dependency management, documentation generation, and collaboration practices in notebooks \cite{Krxa,1JJ8,IUr7}. There are also many solutions for ML-related components such as data validation \cite{pxhw,xpTX}, training data creation \cite{wCk6,enBo}, model building \cite{KcIo,6nq1}, and fairness assessment \cite{zE76,PTwM}. However, without access to study ML products, it is challenging to design and rigorously evaluate interventions at the product level (incl. architecture, collaboration, and documentation).

\textbf{Open-source datasets exist for ML projects, but those are not representative of \emph{ML products,} and not suitable for answering research questions related to ML product development.} Several papers introduce datasets of ML \emph{projects} on GitHub (not necessarily ML \emph{products}). For instance, Gonzalez et al. \cite{YDPU} collected over 4500 ``applied AI and ML'' repositories. The quality of this dataset was criticized for inclusion of toy projects by Rzig et al. \cite{Ku84}, who then curated a new dataset with 2915 ML projects for studying the adoption of continuous integration (CI) practices. Similar datasets of ML projects were curated by van Oort et al. \cite{EMW5} and Tang et al. \cite{wI1Q} for purposes such as discovering the prevalence of code smells and refactoring practices in ML projects. Widyasari et al. \cite{4LDy} shared a manually-labeled dataset of 572 ``engineered ML projects,'' referring to the need for a higher-quality dataset that should include engineering practices. While all these datasets can provide various insights about \emph{ML} \emph{projects}, they are not promising for studying concerns related to developing ML \emph{products} (we analyzed all of them to find only four projects that we consider as products). We argue that insights about architecture \cite{OmEh}, technical debt \cite{TUkr}, and differences in addressing ML and non-ML issue reports \cite{dx7s} from studying ML projects might not well generalize to ML products.

Despite curating many datasets of ML-related projects presented above, researchers have not succeeded in finding ML products in open source, apart from one frequently highlighted example, FaceSwap \cite{1n0g}. Closest to our work, Wan et al. curate a dataset of ML products where a model is used in the context of some non-ML code (in their case to test the code \emph{using} the model). Unfortunately, all their projects are toy projects, such as a 0-starred fire-alarm ``application'' with three commits that uses an object detection model to get object labels from an input photo and prints ``alarm'' if it detects the keyword ``fire'' in the label \cite{N5c4}. 

Existing datasets of ML projects and toy examples limit the generalizability of research to real-world industry-style \emph{ML} \emph{products}. Even worse, if a reader is not aware of the types of projects in a dataset (which is not obvious when the example provided is FaceSwap), it is easy to incorrectly generalize findings to \emph{ML products}. Thus, there is value in a dataset specifically dedicated to ML products, which motivates us to curate and study such a dataset.

\section{Contribution A: Curating the Dataset}\label{h.ephbsc4lhudp}
\label{methContA}

Identifying open-source ML products was surprisingly difficult and turned into a research project in itself. Searching with keywords like ``machine learning'' in READMEs, as in prior work collecting open-source ML projects \cite{YDPU}, does not work because (a) the vast number of ML projects (libraries, notebooks, research experiments, demos) is much larger than the much smaller number of ML products and (b) ML products do not always explicitly advertise their use of machine learning, especially when used for smaller optional features. For example, only one of the top 500 search results on GitHub for ``machine learning'' is an ML product and 13 of our 30 analyzed ML products do not mention machine learning in their README, such as the video-conferencing application (P29) which uses facial expression detection as an add-on.

Instead, we explored and iteratively refined a new search strategy combining domain knowledge, code search, and manual analysis in a process that is specifically designed to scale to search across all of GitHub. In a nutshell, our approach is based on the following insights:

\begin{compactitem}
	\item Targeting end users, ML products have a user interface (mobile, web, desktop, command line), whereas most other ML projects do not. We rely on code search to identify code relating to user interfaces.

	\item ML is used in products usually through a small number of libraries and APIs, whether to train a custom model, to load a serialized model, or to call a remote API service. We rely on code search to identify the use of ML in implementations.

	\item The final distinction between ML products and ML projects requires human judgment (all our attempts at automation yielded poor accuracy). We develop heuristics to prioritize which projects to analyze to manage scarce resources for manual analysis.

	\item Code search at the scale of GitHub is challenging. We carefully design a multi-step pipeline that incrementally reduces the search space, eliminating many projects that are not ML products with cheaper analyses before more expensive analysis steps are required.

\end{compactitem}

Each insight makes assumptions that enable the search to scale and find relevant ML products, but each assumption may lose some ML products that do not meet them, such as user interface mechanisms not captured (e.g., game engines) and models not detected (e.g., custom k-nn implementations). Our approach cannot ensure finding an exhaustive list of all ML products – it is a best-effort attempt to collect as many ML products as possible with reasonable resources, in the face of a very difficult search challenge (see limitations below). 

\subsection{Search Space and Scope}\label{h.lhjj3kr56gut}
We search for ML products on GitHub. GitHub is by far the most popular platform for open-source projects, whereas more specialized platforms such as Hugging Face only host ML models. We only include popular project repositories (over 50 stars) that have been maintained recently (updated after 2019-01-01), and that are documented in English – constraints that are common in open-source research. We restrict our analysis to desktop and web applications written in Javascript, Python, Java, and C\# (most popular languages for such applications \cite{8yZB,IeKU}) in addition to mobile apps for Android and iOS.

\subsection{Search Pipeline}\label{h.74dmwwujr61i}
To scale the search, we proceed in five steps, with increasing per-project analysis cost in each step.

\subsubsection{1. API search} We start with a very scalable step to retrieve a vast overapproximation of candidate projects with the GitHub Search API. We retrieve all GitHub repositories using any of the four programming languages as the primary language and all repositories matching the keywords ``android'' or ``ios.'' We additionally restrict the search to stars and commit date, as mentioned above. Where necessary, we partition the search space by date to overcome GitHub's maximum of 1000 search results. At this stage, we identified 430,902 candidate repositories (cf. Table~\ref{numbers}).

\tableA

\subsubsection{2. Metadata and README filter} We retrieve each candidate project's README and GitHub metadata (including ``about'' description and tagged topics) through the GitHub API. We exclude obvious non-product repositories by matching keywords such as ``framework,'' ``tutorial,'' and ``demo'' in the description or README. In line with similar efforts, we remove archived and deprecated repositories (e.g., keywords ``deprecated'' or ``obsolete''), forks, and repositories with non-English descriptions (using an off-the-shelf model \cite{tBF5}). We report the exact filters in the appendix \cite{LwCB}. We manually validated a random set of 100 filtered projects finding no incorrectly filtered projects. A total of 300,508 repositories remained after applying this filter.

\subsubsection{3. Product filter} To detect user interfaces, we rely on code search, performed locally after cloning each candidate repository. We curated a list of code fragments indicative of 130 common frameworks for user interfaces, such as ``com.android.application'' in a gradle.build file for Android mobile applications, ``import javax.swing'' for Java desktop applications, and ``from flask import Flask'' for web applications in Python (see appendix for details \cite{LwCB}). We remove repositories that do not contain any of these code fragments, leaving us with 26,386 potential products for further analysis.

\subsubsection{4. ML filter} To identify the use of machine learning, we again rely on code search, based on curated lists of code fragments indicative of ML libraries and APIs. We count occurrences of calls to any of 99 ML libraries or APIs (e.g., ``import caffe'') and of serialized models (e.g., files with .tflite, and .mlmodel extensions). In addition, as a noisy last resort, we count occurrences of 20 ML keywords, such as ``machine learning'' and ``NLP'' in any source or text files (including comments and documentation) to catch less common libraries and custom implementations. At this point, 11,257 projects pass at least one ML-related filter.

\subsubsection{5. Manual inspection} The final step with by far the highest per-repository cost is to manually validate whether a repository is an ML product. One or more authors with extensive expertise in ML products inspected the repository, its description, and (when needed) its code to judge whether the repository is indeed an ML product –  this typically took 30 seconds to 20 minutes per repository. Our definition of ML product in Sec.~2 is the result of multiple iterations and refinement, for example, establishing requirements for purpose and documentation, for which we discussed 272 early-inspected projects as a group (of which we considered 94 to be ML products) to arrive at a stable definition which provided us with a high inter-rater reliability (n=40, kappa=0.77). A few repositories near the decision boundary were discussed by all authors until a unanimous consensus was reached. We inspected about 4,000 of the 11,257 remaining repositories, prioritizing our resources based on match counts for our ML filters, stratified product category, language, and ML filter. In each strata, we stopped when we reached 30 consecutive false positive repositories, for example after inspecting 216 Android mobile apps.

\subsection{Limitations and Threats to Validity}\label{h.8peznfj26041}
To make the search feasible we had to make various compromises, arriving at the described design. Given the various heuristics, our approach represents a best-effort attempt and cannot claim producing an exhaustive or complete list of ML products. As discussed, we may have missed ML products in other languages, using other GUI frameworks, or less common ML libraries. Additionally, our approach involved manual inspection, which, despite best efforts, opens the possibility of human error and subjectivity.

Our search heuristics prioritize false positives over false negatives, and we designed our approach accepting low precision (discarding many repositories in the last manual validation step) to ensure high recall. While we would have preferred to formally evaluate recall (i.e.,  whether we missed any ML products) by comparing our dataset against any existing ground-truth dataset of ML products, such a dataset does not exist. As a substitute, we attempted to collect ML products independently by seeking input from industry practitioners through platforms like Quora, Reddit, LinkedIn, Twitter, and a 32k-member Slack channel in the field of data science; but aside from numerous replies expressing interest in our dataset, we only received suggestions for two repositories, both of which we determined not to be ML products according to our definition. Additionally, we compared our dataset to other existing datasets of ML projects \cite{YDPU,4LDy,TBW6,wI1Q} but did not find any additional repositories that satisfy our definition of ML products in those datasets. In fact, those datasets only contained a total of four ML products, all of which we detected in our dataset. While all this raises our confidence, we cannot formally assess recall.

\subsection{The Open-Source ML Product Dataset}\label{h.rcen2wim2svk}
\label{findingcontA}

In total, we found 262\footnote{Removing duplicates, such as same repositories for Android and iOS apps.} ML products (cf. Table \ref{numbers}, full dataset in appendix \cite{LwCB}). The average ML product in our corpus has 1495 stars, 28 contributors, and is 325MB in size. Over half of the ML products are written in Python; most are web applications.

The dataset comprises a diverse range of products, some of which have a significantly larger number of users and a more professional look than others. For instance, Seek (P2 in Table \ref{samples}) is a mobile app for identifying plants and animals using image recognition, downloaded over 1 million times and reviewed by over 38k users with robust support from the established iNaturalist community, who maintains a dedicated website and continuously improves and maintains the app. In contrast, NotionAI MyMind (P5), an Android app developed by a single contributor, uses an ML classifier to automatically tag images and articles, with a simple user interface, rare updates, and under 5,000 downloads from the Play Store. Approximately half of the products in our dataset have a professional presentation like Seek; those generally have more stars and a larger codebase. The others seem to be personal-interest projects released as a product.

\section{Contribution B: Learning from the Dataset}\label{h.p3f07bfp80w2}
We created this dataset due to limited access to industry products, which hinders research advancements and education in the field. Now that we have this dataset, we can finally attempt to answer numerous research questions about ML products that have accumulated in many past studies with open-source products, where previously we had to rely on interviews or experience reports from industry practitioners. Given the breadth of topics, we cannot cover everything in a single study. In this paper, we explore a wide range of topics rather than going into depth on a single one, to \emph{contribute new knowledge} and achieve two secondary objectives: \textbf{(1) }\emph{Characterize the dataset:} Our analysis with broad research questions will implicitly characterize the products in our dataset and enable other researchers to effectively use it and interpret derived findings (cf. limitations of existing work in Section \ref{relWork}) -- this also helps to explore how similar the ML products in our dataset are to ML products described in interviews and experience reports. \textbf{(2)} \emph{Identify when deeper analysis of the dataset is feasible and promising:} Our research questions from different topics, such as collaboration, architecture, and development process will identify what kind of questions are worth going into deeper before committing resources to in-depth analyses for individual research questions.

\subsection{Research Method}\label{h.lvabn97dc0l6}
\label{methContB}

First, we curated a list of research questions relevant to the study and designed qualitative and quantitative strategies to answer them. Given the novelty of the dataset and questions, we heavily rely on qualitative analysis involving substantial manual effort. Therefore, we decided it would be more manageable to analyze a sample of 30 products from the dataset.

\subsubsection{Deriving Research Questions}\label{h.ysk7954m31n5}
We selected research questions that are not only of interest to the research community but can also be feasibly answered by analyzing open-source ML products (e.g., we did not find artifacts that would allow us to answer \emph{``How do data scientists elicit, document, and analyze requirements for ML systems?''}). For this selection, we employed a two-step process. 

First, we explored the existing literature to identify topics that are of interest to researchers in the field, such as the challenges faced by practitioners building ML products, collected in our recent meta-survey of interviews and surveys \cite{qMjN}. We identified numerous topics of interest, such as collaboration, architecture, process, quality assurance, MLOps, and responsible AI.

Second, we examined our dataset of ML products to identify potential research questions (RQs) that could plausibly be answered with open-source data. We explored 15 randomly selected ML products qualitatively, taking multiple pages of notes for each product. We immersed ourselves in the source code, documentation, contributor profiles, issues, and any other available information provided on associated websites; we identified the ML and non-ML components to familiarize us with common structures. This gave us a sense of what kind of questions can be reasonably answered.

In the end, we selected six questions spanning the entire life cycle: \(\bullet\) \textbf{RQ \#1 (Collaboration):} How interdisciplinary are open-source ML product teams and how do they divide their work? \cite{rrC2w,y2A6} \(\bullet\) \textbf{RQ \#2 (Architecture):} How are open-source ML products architected to incorporate models? \cite{Nrd9,OmEh} \(\bullet\) \textbf{RQ \#3 (Process):} What model-product development trajectory do open-source ML products follow? \cite{gHwA,kLg8} \(\bullet\) \textbf{RQ \#4 (Testing):} What and how are the open-source ML products and their parts tested? \cite{fC0u,3rn7} \(\bullet\) \textbf{RQ \#5 (Operations):} How are open-source ML products designed for operation? \cite{DipR,SuxN} \(\bullet\) \textbf{RQ \#6 (Responsible AI):} What responsible AI practices are used in open-source ML products? \cite{8V2w,HCDa}

While each identified topic could warrant a dedicated study and deeper analysis, here we provide initial answers for each and explore opportunities for future research.

\subsubsection{Analysis and Synthesis}\label{h.8h1gfh1zuyz7}
Without existing established measures, we found a manual, mostly-qualitative analysis to be more responsive and effective than a narrow quantitative analysis at scale \cite{GkY9,9Iu0}. To find answers to the RQs, the ML products required an in-depth examination of the code, analysis of contributor activities, and thorough inspection of related documents. While we automated some measures, such as contributor percentages and modularity scores, designing them also required initial manual investigation and manual classification of ML components. This made the process quite labor-intensive, requiring 10-15 hours per product. Consequently, we analyzed a sample rather than every product in the dataset.

\emph{Sampling:} We analyzed the 30 products shown in Table \ref{samples} (11.5\% of the dataset), which was manageable for our manual analysis. Rather than attempting statistical generalizations, our goal was to gain rich insights from the dataset. Thus, we aimed to sample a diverse range of products using case-study research logic \cite{oMJD,7NRz}. We used \emph{information-oriented selection} to select popular and large products (\emph{extreme/deviant cases}), because we expect them have a richer history to analyze, and we included a random selection of other products for the \emph{average cases}. To represent different kinds of products, we stratified selection across the three genres: mobile, desktop, and web. Specifically, we select two products with the most stars per genre (P8, P9, P16, P17, P21, P28), two products with the most contributors per genre (P3, P6, P14, P19, P24, P25), the product with the largest code size per genre (P1, P11, P29), and five randomly selected products per genre. The sample set of ML Products vary in terms of stars (avg. 5k, min. 63, max. 42k), no of contributors (avg. 82, min. 1, max. 374), and LOC (avg. 564k, min. 1106, max. 2745k), as shown in Table \ref{samples} and appendix \cite{LwCB}.

\emph{Analyzing Products and Card Sorting:} We conducted a comprehensive qualitative analysis of the sampled products. We describe the specific analysis steps separately for each research question below, but generally, we follow the strategy: two researchers carefully examined the GitHub repositories, addressing each research question individually for each product, involving tasks, such as reading documentation, identifying ML and non-ML components in the source code, measuring modularity, examining contributor profiles, analyzing commit history, and reviewing issues. The entire research team regularly met to discuss, clarify understanding and resolve disagreements, and organize findings. To organize and find patterns among the products, we performed card sorting \cite{ZAks}. Each product was represented by a card for each RQ, describing our findings for that particular RQ, and we iteratively grouped these cards to identify patterns within RQs. Additionally, we searched for associations across patterns from different RQs. We share analysis artifacts, including a Miro board and spreadsheets, in the appendix \cite{LwCB}.

\subsubsection{Threats to Credibility and Validity}\label{h.huw07591kc1v}
Despite following best practices for qualitative research as we discussed, this part of the research shares common threats encountered in qualitative research \cite{GkY9,9Iu0}. Given the small sample size and sampling strategy, statistical generalization is not suitable and not advised. Readers should compare this to a study with 30 interviews. While we followed standard practices for coding and memoing during the analysis of the products, we cannot completely eliminate biases introduced by the researchers. In addition, we only access public information and do not have access to offline activities -- hence, our findings should be interpreted accordingly. 

\subsection*{RQ \#1: How interdisciplinary are open-source ML product teams and how do they divide work?}

Interdisciplinary collaboration is difficult, as has also been found in building ML products \cite{y2A6,rrC2w}. The transparency of open-source development allows researchers to study many aspects of collaboration, as demonstrated by numerous past studies on team collaboration, pull requests, and diversity, e.g., \cite{iYmX,G2dk,B8Ur}. To understand \emph{interdisciplinary} collaboration in open-source ML products, we explore team composition in terms of the number of contributors and their backgrounds, who works on ML and non-ML code of the product, and how tangled ML and non-ML code are in terms of co-changes. The findings can help study existing challenges such as siloed development and guide further studies in collaboration.

\subsubsection{Method}\label{h.utv3rho5a54m}
To analyze contributor backgrounds and numbers (\emph{Findings 1-2}), we collected contribution data from GitHub and identified the \emph{core} contributors as those collectively responsible for 80\% of all commits (in line with past work \cite{RJSl,Jz8a}). We manually classified each contributor's background as \emph{SE-focused,} \emph{ML-focused}, or \emph{other} (e.g., physics, finance), based on public self-description, professional title, and education history as found on their GitHub profile, LinkedIn profile, and personal, project, or company websites, if available. If the classification was not obvious (e.g., because of limited public information) we classify their background as \emph{``unsure.''} 

To study how developers from SE and ML backgrounds contribute to ML and non-ML code (\emph{Findings 3-4}), we separated the ML and non-ML code after excluding documentation and binaries: We manually categorized code associated with model training, prediction, and pipeline as ML-related, while all other software infrastructure and graphical user interface (GUI) code fell under non-ML, typically at the granularity of files. Finally, we automatically analyze the commit history to attribute code changes to contributors.

To analyze the coupling of ML and non-ML code (\emph{Finding 5}), we analyze co-edits in the product history, known as \emph{logical coupling} \cite{Di9F,iqpc}. Specifically, we compute a \emph{relative coupling index} that indicates whether the ML and non-ML parts are more or less coupled than would be expected if all changes were randomly distributed across files. A low \emph{relative coupling index} indicates that changes are typically isolated to only ML code or only non-ML code, whereas a high \emph{relative coupling index} indicates that ML and non-ML code are often changed together, signaling low modularity. To compute \emph{relative coupling index}, we compute coupling using \emph{evolutionary coupling index (ECI) \cite{Di9F}} and then divide \emph{ECI} by the probability of coupling of random edits to normalize the effect of size, as ML and non-ML code size differs significantly (the average product has 971k LOC  of non-ML code and  23k LOC of ML code).

\subsubsection{Finding 1: Most of the core contributors are software engineers (74/92). Finding 2: Many products have a single core contributor (13/30)}\label{h.1dw11vpecc1f}
In our sample of 30 products, we identified 140 core contributors, among whom we could classify 92. Among the 92, we found 74 contributors as \emph{SE-focused and} 10 contributors (spread across eight products) as \emph{ML-focused}. \figureCont For 16 products where we could classify all core contributors, many (9/16) had exclusively \emph{SE-focused} contributors. We found a single contributor (in P24) who self-identified to be an expert in both SE and ML.

\subsubsection{Finding 3: There is little evidence of clear silos, with core contributors commonly committing to both ML and non-ML code, regardless of background. Finding 4: Team responsibilities are rarely assigned and recognizable in the commit history}\label{h.tg3x34ahq11n}
In contrast to the  \figureContCode widely reported problem of \emph{siloing} in industry teams \cite{rrC2w}, we do not find a clear delineation by background of who contributes to the ML and non-ML code. We often find contributors of either background working on both parts.\\

Only five products (P13, P21, P24, P25, P29) publicly documented team structures with assigned roles and responsibilities for team members, but even then the assigned responsibilities do not always reflect the commit activities. For instance, even though P29 has an explicit data team, we do not find commits from the data team to the ML repository, but \emph{SE-focused} contributors change both the ML and non-ML code. We conjecture that some offline collaboration is not visible in the open-source repository.

\figureModularity

\subsubsection{Finding 5: ML and non-ML code are often changed together, indicating low levels of modularity}\label{h.rrj3x72010zd}
While models are usually assumed to be modular components in a system, our analysis reveals that many products (12/30) in our sample exhibit frequent co-changes of ML and non-ML code. For example, P27 has a very high \emph{relative coupling index} between ML and non-ML code – we found that this product has a custom script for training a speech recognition model, where the ML code directly updates the user interface (UI) button based on the prediction result; any modification to the UI properties requires an update to the model script to accommodate the change, causing frequent co-changes of the UI files and the model script. Conversely, although we indeed found many products (18/30) that have coupling lower than random, it is not as low as could be expected from fully modular components of a product. 

\subsubsection{Discussion (open source versus industry): Open source ML products mirror the startup style of development more than big tech projects}\label{h.bzwpmai8l70w}
In line with general trends in open source \cite{OXc0,pYnK}, we find relatively small teams developing ML products in our sample, in contrast to often large teams reported to build ML products in big tech companies \cite{swt2,0qnc}. Our open-source ML products are closer to the average team size of 2-5 members in startups \cite{dUhQW}. While prior studies report misaligned responsibilities that do not reflect developers' abilities or preferences across all kinds of organizations building ML products \cite{y2A6,FzPp,rrC2w}, the fluent and broad responsibilities and collective code-ownership resemble characteristics commonly seen in startups. Overall, the open-source ML products seem more reflective of activities in the vast majority of ML projects outside of big tech organizations, that have their own distinct and often understudied challenges \cite{R1NT,rrC2w,RmKB}.

\subsubsection{Implication 1: Researchers should study the challenges of teams that do not have access to data scientists and explore providing assistance}\label{h.4zrms52rsx9r}
Most studies on ML development focus on the challenges of data scientists, perceived as the dominant or novel role. Open-source ML products seem to be dominated by software engineers, who adopt ML tools, often with limited apparent participation from data scientists. Past interview studies have already established pitfalls of software engineers adopting ML without explicit training \cite{rrC2w}, such as inadequate feature engineering and insufficient evaluation. Open-source ML products provide an opportunity to study such problems in public artifacts. In addition, ML products developed without dedicated data scientists are likely common also in industry, outside of big tech, and likely to become more common as data science becomes more accessible (e.g., with AutoML and prompt engineering) – researchers should explore how to support software engineers with limited data-science expertise in building ML products responsibly, for example, through analysis, automation, and smart assistants. Recent tools for detecting data smells \cite{pLeD} and data leakage \cite{lske} and for anticipating fairness issues \cite{9g7T} provide encouraging starting points. Conversely, a few ML products in our sample are developed by data scientists without software engineers – such cases are better researched \cite{rrC2w,Jd8i}, but can equally benefit from further studies and support.

\subsubsection{Implication 2: Researchers should investigate sources of non-modularity and develop tools and guidance}\label{h.yhn6qqjyu1rk}
While interactions among multiple models are a well-known problem (\emph{``changing anything changes everything'')} \cite{t8sQ,94yr,Ijae}, models are usually considered natural modules in a software design with clear and simple interfaces \cite{RfQCh,ftJI}. Yet, we found surprisingly frequent co-changes of ML and non-ML code. Research should explore the sources of this non-modularity and have a unique opportunity to do so with our dataset. Research should identify or create design strategies to isolate change, possibly coded as design patterns (current ML-related design patterns rarely consider the interaction of ML and non-ML code \cite{zLpk,V97G}), to guide practitioners toward more modular designs. Positive examples in the dataset could serve as illustrations in educational materials.


\subsection*{RQ \#2: How are open-source ML products architected to incorporate models?}

Researchers have highlighted how ML can influence the architecture of software products \cite{swt2,Nrd9}. To comprehensively understand the product structures and the incorporated models, we explore architecturally relevant aspects such as model type, usage, importance, integration of multiple models, pipeline automation, documentation, and big data infrastructure.

\subsubsection{Method}\label{h.emt3crto7ly2}
To understand the overall structure of the model and product, we conducted a comprehensive manual analysis of the ML and non-ML code in the repositories. We analyzed the code with a focus on the following artifacts: code structure and data flow as it pertains to models -- identifying how the models are created, where and how they are called, and how the model predictions are processed and used. We also reviewed their documentation, relevant blogs and forums, associated web pages, and related repositories under the same personal or organizational account. We then sorted our findings and grouped those into categories, using card sorting techniques \cite{ZAks}, guidance from previous research, and domain knowledge from our research team.

\subsubsection{Finding 6: About half of the products rely exclusively on third-party ML models (13/30)}\label{h.8c9ficn6gn34}
We identified 15 products that use \emph{third-party models} via libraries (e.g., Tesseract OCR), external APIs (e.g., ClarifyAI), or load pre-trained model files from a remote repository. In contrast, 17 products \emph{self-train models}. Two products use both \emph{third-party} or \emph{self-trained models} (P12, P30). For instance, the optical music recognition application, P12, uses a self-trained model to classify music symbols and an existing OCR library for classifying text. 

\subsubsection{Finding 7: The importance of the ML models to the product varies, with about half using them as optional functionality only (13/30)}\label{h.uqy9e47c8fux}
We found the importance of ML models to vary considerably across different products. The model is the \emph{core} functionality in 11 products, as there would be no product without the model (e.g., the OCR model in the OCR scanner app P1). There are 6 products that may still provide value without the model, but the model is a \emph{significant} functionality (e.g., the OCR model in video subtitle editor P15 that could potentially operate on manual inputs). In 13 products, the model provides \emph{optional} functionality, serving as a nice-to-have add-on (e.g., facial expression recognition in video conferencing app P28). Whether a \emph{third-party} model is used (\emph{Finding 6}) is not necessarily associated with the model's importance: We found products investing substantial effort in self-trained models for \emph{optional} functionality (e.g., P6, P26, P29) and products relying on a third-party model for \emph{core} functionality (e.g., P1, P12, P13). Products with models as \emph{core} tend to be smaller and have fewer contributors (avg. 112k LOC, 1.4 contributors), while \emph{optional} models are often added to larger products with more contributors (avg. 754k LOC,  8.1 contributors).

\subsubsection{Finding 8: Automation using model predictions is uncommon (5/30), with most products keeping humans in the loop}\label{h.w6pxwder3lim}
A central question in human-AI design is how to use or present model predictions and whether and how to keep humans in the loop \cite{DeO4,ftJI}. We find only five products that use model predictions to fully \emph{automate} actions (e.g., keyword spotting app P27 executes gameplay instructions based on recognized voice commands). Two products \emph{prompt} users to confirm an action (e.g., deepfake software P17 asks for confirmations on image previews between each processing stage). Most products (23) in our sample merely \emph{display} predictions,  leaving decisions about actions entirely to users (e.g., trading app P21 graphically presents investment predictions).

\subsubsection{Finding 9: Most products use raw model predictions without any post-processing (21/30). Finding 10: Products that automate actions are more likely to further process model predictions}\label{h.s87s20fgw3ko}
Products may check, process, and integrate model predictions in many ways, some now encoded as patterns, such as \emph{two-phased predictions} for resilient serving of models \cite{zLpk,YIKzr}. However, most analyzed products (21/30) trust model predictions and display them without any further processing. Only two products incorporate additional architectural tactics around model predictions: Plant identifier app P2 uses a two-phase prediction system, combining a local model with an online model for low-confidence cases (known as \emph{two-phase prediction pattern} \cite{zLpk}) and subtitle editor P15 performs extensive checks on texts predicted via optical character recognition and language translation before presenting them. In addition, three products incorporate a confidence score threshold to filter low confidence predictions (P9, P27, P29) and another three offer a retraining option for the model if performance proves unsatisfactory (P5, P12, P28). Interestingly, P11 uses machine learning to check the results from a non-ML API. The few products that automated decisions based on model predictions (Finding 8) process predictions further, by offering retraining mechanisms (3/5) and confidence checks (1/5).

\figureSankeyA

\subsubsection{Finding 11: Many products  use multiple models (18/30), though those models are mostly independent (11/18)}\label{h.uoi0zdr8e9fh}
Interactions among multiple models is a frequently raised challenge in industry, where a minor change in one model can trigger cascading changes across the product \cite{Ijae,94yr,t8sQ,ftJI}. While 18 products use multiple models, those perform independent tasks in 11 products\footnote{Numbers do not add up, as some products fall in multiple categories.}: 7 products use models for \emph{separate functions} unrelated to each other (e.g., P15 uses one model for OCR and another for speech-to-text) and 7 products provide \emph{alternative} models for the same function (e.g., P26 provides a choice between two clustering models). Five products \emph{sequentially compose} models \cite{ftJI} (e.g., P12 passes text recognized by an OCR model to an entity recognition model). Two products use models for \emph{collective decision-making} (e.g., P9 combines multiple classifiers to generate personalized news feeds).

\subsubsection{Finding 12: Pipeline automation is not common in open-source ML products}\label{h.1yzpu86at7cz}
Switching from a static mindset and notebooks to pipeline automation is a commonly reported challenge \cite{e6bb,rrC2w}. Among the 17 products that use self-trained models (\emph{Finding 7}), training is often not automated.  We did not find any model training pipeline for four products (P2, P18, P22, P30; we cannot tell if training happens offline or in a private repository). Four products (P6, P10, P12, P16) require manual execution of sequential training steps; two products automate only data retrieval (P11, P24)\footnotemark[\value{footnote}]. Four products (P9, P15, P27, P29) have GUI-integrated training pipelines that can be separately activated via GUI actions. Only four products (P14, P21, P23, P26) feature fully automated training pipelines to consistently fetch the latest data and deploy updated models. 

\subsubsection{Finding 13: We do not find much effort on data or model documentation}\label{h.6yezq0rk4en5}
Both industry and academia view model and data documentation as important for, among others, collaboration, accountability, and reuse \cite{tPzV,Fhfs,L6sa,agC2}, but adoption in industry is rare and perceived as challenging \cite{tPzV,Krxa}. In our sample, the 17 products using self-trained models (\emph{Finding 7})  provided minimal and mostly scattered documentation for models and data, if any. Regarding \textit{model documentation}, only one product, P29, provides high-quality model documentation (in the form of a model card \cite{9y7Z}). Other products have at most brief instructions for using the model API or descriptions of the model architecture for the data scientists. \textit{Data documentation} was mostly limited to presenting a data schema, occasionally mentioning the volume of the training data, or simply providing a link to their data sources. We found no use of datasheets \cite{EggD} or similar templates. 

\subsubsection{Finding 14: Most products do not use big data infrastructure (23/30)}\label{h.beowx8mtu3tu}
Scalability is reported as a common, important architectural challenge for ML products, for handling large datasets and distributing expensive training and inference jobs, resulting in frequent reliance on big data infrastructure. However, we did not find the use of local or self-hosted big data infrastructure (such as Hadoop and Spark) in any of our sample ML products. Seven products contain code related to cloud services for storage, computing, monitoring, and search (e.g., Amazon S3, EC2, CloudWatch, and Elastic Cloud). 

\subsubsection{Discussion (open source versus industry): Open-source ML products share many of the development decisions and challenges discussed in industry studies}\label{h.bpuh46130oil}
In line with the industry trends, many open-source projects leverage third-party models rather than building models from scratch, a pragmatic choice given the cost, time, and skill requirements involved in developing models with limited resources \cite{KKry,uE8C,R1NT}. We find models used for various tasks with varying levels of importance in a product, reflecting the diversity of the ML products. Open-source ML products with models as \emph{core} more resemble startup-style projects, whereas attempts to integrate models as enhancements to existing products are found throughout the industry, including many established corporations. While there are large variations within our dataset, open-source ML products tend to lean toward the less complex end of reported ML products in the industry, with simpler architectures with few models, limited automation, and less need for massive scale. The lack of pipeline thinking, overly trusting model predictions, and poor documentation mirror practices repeatedly criticized in industry projects \cite{DipR,tPzV,qMjN}, and while the more experienced and well-resourced companies work toward better practices \cite{3rn7,9y7Z,zLpk,0qnc} those challenges are still common in many newer and smaller organizations.

\subsubsection{Implication 3: Researchers should study patterns and test interventions for different architectural choices}\label{h.yx6thee1nnlq}
While researchers and practitioners argue for the need to implement safeguards \cite{wMcm,fesc}, prepare for the evolution of third-party models \cite{XOW6,KKry}, design effective and safe human-AI interaction models \cite{bNoW}, and integrate multiple models \cite{Ijae,t8sQ}, researchers rarely have access to enough products to detect patterns and validate solutions on a range of systems. Even if some of the practices are rare in open source, our dataset has many and diverse projects to study existing patterns and to provide a testbed to evaluate the consequences of different design interventions, such as design patterns to isolate models or data-flow analyses to track whether and how multiple models in a product interact. It also provides opportunities for deeper investigation in certain aspects, such as employing firehouse studies \cite{lcTM} to interview developers when certain events occur.  Additionally, it allows researchers to conduct longitudinal studies to understand the evolution of the team and the architecture over time. We have not seen any prior longitudinal studies of projects in this field (common in MSR-style research), likely due to a lack of access.

\subsubsection{Implication 4: Educators should use open source to develop teaching materials}\label{h.ba22pwefto5n}
In a field where access to concrete implementations is scarce and educators often rely on demo projects or second-hand reports, open-source ML products can be a valuable educational resource to showcase system design strategies and challenges, whether as illustrations in lectures and blog posts, as foundations for homework assignments, or as in-depth case studies as in the \emph{Architecture of Open Source Application} books \cite{Hr7T,GF2x}. The dataset has sufficient variety to cover simpler projects suitable for beginners, as well as sophisticated products built by large teams to study architectural design decisions.

\subsubsection{Implication 5: Companies, foundations, and governments should explore strategies to sustain model and big data infrastructure}\label{h.yuxi99qx6gl9}
Unlike revenue-generating commercial products, most open-source ML products in our sample (27/30) did not seek to monetize their products. The potential high recurring cost for model APIs and cloud computing may prevent open-source developers from scaling their products or from building certain products in the first place. Some open-source products may find a path to secure funding; for example, Seek by iNaturalist P2 was supported by various nonprofit foundations before establishing its own nonprofit with a seed grant. While companies, foundations, and governments often support open source (e.g., free hosting and CI on GitHub; Sovereign Tech Fund, NSF POSE), sustained support for model APIs and cloud computing is less common. Such support seems essential to encourage open-source innovations as alternatives to commercially dominated ML products.

\subsection*{RQ \#3-6: Process, Testing, Operations, and Responsible AI}

Due to the page limit, we only report brief findings from the remaining four research questions, but refer the interested reader to our appendix \cite{LwCB} for details on the methods, findings, and discussions. The following findings 15-16 relate to RQ \#3, finding 17 to RQ \#4, findings 18-20 to RQ \#5, and finding 21 to RQ \#6. 

\subsubsection{Finding 15: \emph{Product-first development} (16/30) is more common than \emph{model-first} (7/30). Finding 16: When the model is the \emph{core} functionality, it is always developed first}\label{h.24ka0ctek0qy}
In prior work, we found that some projects start with models and later \figureSankeyB build products around them (model first) whereas others adopt a product-first approach – each creating distinct challenges \cite{rrC2w}. In our open-source sample, we observe a greater prevalence of the product-first trajectory, which may be attributed to most contributors being software engineers (\emph{Finding 2}) and many products adding machine learning for optional functionality to existing products (\emph{Finding 7}). Noticeably, products with models as \emph{core} are always developed \emph{model-first}. For example, deepfakes software P17, created the model first and added a GUI a year later to make the model accessible to end users. 

\subsubsection{Finding 17: Testing regular software functionality is common (23/30), model testing is notably scarce (8/30), and data validation is rare (2/30)}\label{h.jdyvt0veoa34}
Standard software testing practices are widespread, but model evaluation is less prevalent in our sample. Even among the eight products that included model evaluation scripts, three (P3, P9, P29) approached model testing like unit testing, asserting that predictions match expected values. The rare cases (P6, P21) in which data validation is conducted involve only minimal checks for schema and value ranges.

\subsubsection{Finding 18: Only a few products (8/17) have mechanisms for evolving models}\label{h.kb5v89wgpftg}
Of the 17 products with self-trained models, five products (P9, P16, P23, P27, P30) offer users the option to retrain products at run time and three products (P14, P21, P26) continually retrain their models by fetching up-to-date data from their data sources. This aligns with our recent finding that many product teams have a static view of models \cite{rrC2w}.\looseness=-1

\subsubsection{Finding 19: Model monitoring is almost non-existent (1/30)}\label{h.pao5zgfpfsu7}
Despite the heavy emphasis on observability in industry and academic literature for detecting failures and degradation \cite{SuxN,qBTJ}, 29 out of the 30 products do not collect telemetry and have no monitoring infrastructure. Only P21 incorporated telemetry for financial forecasting. In addition, P9 uses Amazon CloudWatch, but not for model monitoring.

\subsubsection{Finding 20: MLOps tools are not used}\label{h.c2kmrkspz049}
We did not find use of any popular MLOps tools for tasks such as automating deployment, testing, monitoring, and data cataloging, in any of the products. Given that many of these products do not incorporate retraining mechanisms (\emph{Finding 18}), they may have less need for MLOps automation. 

\subsubsection{Finding 21: Responsible AI practices (e.g., fairness, safety, security) are not apparent in open-source ML products}\label{h.2my63y92b1ke}
Despite significant attention in academia, we do not find adoption of any responsible AI practices in our sample. Only one product, P17, discusses ethical usage in their \emph{README} \cite{1n0g}, largely limited to disclaimers. Several products include privacy policies and disclaimers unrelated to ML. One educational product (P29) covers responsible AI practices as a subject.

\subsubsection{Discussion (open source versus industry): In comparison to industry, open source showcases similar and more bad practices associated with both model and product evaluation and maintenance}\label{h.gg3w2f5smdag}
Similar to our observations on architecture, open-source ML products exhibit many of the characteristics criticized in past research, with low adoption of tooling and interventions discussed by more experienced and well-resourced organizations – we find the same low adoption of model evaluation \cite{9njJ}, data validation \cite{XLRw}, monitoring \cite{R1NT,qMjN}, and responsible AI practices \cite{8V2w,Krxa,rrC2w}. The open-source ML products seem to reflect the practices of new organizations and smaller teams more than those of big tech organizations; they are likely reflective of the challenges that new teams will experience, especially teams dominated by software engineers.

\subsubsection{Implication 6: Tool vendors have an opportunity to showcase the benefits of automation tooling}\label{h.cdk9uculny05}
Rather than relying on testimonials and narrowly scoped tutorials, tool vendors such as those of MLOps tools can demonstrate their tools in forks of open-source ML products or can even work with open-source developers to integrate them into their products. For example, while researchers have found some public (mostly very small) projects using the versioning tool DVC \cite{KZfV}, none of them were full ML products showcasing the integration of model and product versioning. Similar to Implication 5, tool vendors can provide resources to support open-source communities.

\subsubsection{Implication 7: Research should explore open-source friendly monitoring approaches}\label{h.1hmd7kena94l}
Observability is a key focus of the MLOps community; many researchers and practitioners argue that the unreliable nature of ML and presence of data drift makes monitoring and testing in production crucial to responsible engineering \cite{fCsC,MsJ6}. We conjecture that the privacy-conscious open-source culture and a lack of centralized infrastructure contribute to the minimal adoption of monitoring among open-source ML products. Researchers should explore privacy-preserving and community-operated monitoring solutions compatible with open-source values, ideally through co-design processes with open-source practitioners.

\section{Conclusion}\label{h.55tvkxlj5q}
We offer a dataset of 262 open-source ML products to facilitate research experiments that can benefit from access to the development history and artifacts of ML products, and report 21 findings and seven implications from six research questions. The dataset is a valuable educational resource for both academics and practitioners in ML product development, offering diverse study materials with both large and small ML products, and it also provides ample research opportunities as described.

\section*{Acknowledgment}
The authors would like to thank Bogdan Vasilescu, Rohan Padhye, and Eunsuk Kang for helping with the framing, and their continuous suggestions and feedback. The author would also like to thank the industry practitioners who responded to the queries about ML products on Quora, Reddit, LinkedIn, Twitter, and Slack.

Kästner's, Nahar's, and Zhang's work was supported in part by the National Science Foundation (\#2131477), Zhou's work was supported in part by the Natural Sciences and Engineering Research Council of Canada (NSERC, RGPIN2021-03538), and Lewis' work was funded and supported by the Department of Defense under Contract No. FA8702-15-D-0002 with Carnegie Mellon University for the operation of the Software Engineering Institute, a federally funded research and development center (DM24-1039).

\balance

\end{document}